\documentclass[12pt,preprint]{aastex}
\usepackage{epsfig}
\usepackage{graphicx}
\usepackage{lineno}
\usepackage[dvips]{color}
\slugcomment{Accepted by ApJ Letters on 10 April 2013}

\shorttitle{Redshift Lower Limit of PKS 1424+240}
\shortauthors{Furniss et al.}

\begin{document}

\title{The Firm Redshift Lower Limit of the Most Distant TeV-Detected Blazar PKS 1424+240}

\author{A. Furniss\altaffilmark{1}, 
D. A. Williams\altaffilmark{1},
C. Danforth\altaffilmark{2},
M. Fumagalli\altaffilmark{3,4,5}, 
J. X. Prochaska\altaffilmark{6},
J. Primack\altaffilmark{1}, 
C. M. Urry\altaffilmark{7}, 
J. Stocke\altaffilmark{2},
A. V. Filippenko\altaffilmark{8},
and
W. Neely\altaffilmark{9}}

\altaffiltext{1}{Santa Cruz Institute of Particle Physics, and Department of Physics, University of California Santa Cruz, 1156 High Street, Santa Cruz, CA 95064, USA.}
\altaffiltext{2}{CASA, Department of Astrophysical and Planetary Sciences, University of Colorado, 389-UCB, Boulder, CO 80309, USA.}
\altaffiltext{3}{Carnegie Observatories, 813 Santa Barbara Street, Pasadena, CA 91101, USA.}
\altaffiltext{4}{Department of Astrophysics, Princeton University, Princeton, NJ 08544-1001, USA.}
\altaffiltext{5}{Hubble Fellow.}
\altaffiltext{6}{Department of Astronomy and Astrophysics, UCO/Lick Observatory, University of California, 1156 High Street, Santa Cruz, CA 95064, USA.}
\altaffiltext{7}{Department of Astronomy, Yale University, New Haven, CT 06520-8120.}
\altaffiltext{8}{Department of Astronomy, University of California, Berkeley, CA 94720-3411.}
\altaffiltext{9}{NF$/$ Observatory, Silver City, NM}

\begin{abstract}
We present the redshift lower limit of $z\ge0.6035$ for the very-high-energy (VHE; $E\ge$100 GeV) emitting blazar PKS\,1424+240 (PG\,1424+240).  This limit is inferred from Lyman $\beta$ and $\gamma$ absorption observed in the far-ultraviolet spectra from the \textit{Hubble Space Telescope/Cosmic Origins Spectrograph}.  No VHE-detected blazar has shown solid spectroscopic evidence of being more distant.   At this distance, VHE observations by VERITAS are shown to sample historically large gamma-ray opacity values at 500 GeV, extending beyond $\tau$=4 for low-level models of the extragalactic background light (EBL) and beyond $\tau=$5 for high-levels.  The majority of the $z=0.6035$ absorption-corrected VHE spectrum appears to exhibit a lower flux than an extrapolation of the contemporaneous LAT power-law fit beyond 100 GeV.  However, the highest energy VERITAS point is the only point showing agreement with this extrapolation, possibly implying the overestimation of the gamma-ray opacity or the onset of an unexpected VHE spectral feature.  A curved log parabola is favored when fitting the full range of gamma-ray data (0.5 to 500 GeV).  While fitting the absorption-corrected VHE data alone results in a harder differential power law than that from the full range, the indices derived using three EBL models are consistent with the physically motivated limit set by Fermi acceleration processes.
\end{abstract}


\section{Introduction}
Blazars are active galaxies with a jet closely aligned with the Earth line of sight \citep{urry}.  These objects are the most commonly detected type of extragalactic source at very high energies (VHE; $E\ge$100 GeV).  The VHE gamma rays which propagate through the intergalactic medium are absorbed by extragalactic background light (EBL) photons via pair production \citep{nikishov,gould}.  The EBL %consists of the accumulated and reprocessed radiation of all starlight produced thus far and 
is difficult to directly measure due to strong foreground sources \citep{hauser}.   

Various methods are used to estimate the density of the EBL.  A selection of EBL models include a semi-analytical model \citep{gilmore2012}, an observationally based model which utilizes observations of \textit{K}-band rest frame galaxy luminosity functions in combination with galaxy spectral energy distribution fractions \citep{dominguez} and a model based on starlight emission and dust re-emission templates \citep{finke}.  Each of these estimate photon densities are above the lower limits set by galaxy counts \citep{werner}, and below the limits set by detection of extragalactic VHE photons (e.g., \cite{aharonian2006}).

The interaction between VHE and EBL photons produces a ``gamma-ray horizon," limiting the distance to which VHE sources can be detected.   3C\,279 is the most distant VHE emitting blazar with a spectroscopically measured redshift.  By spectroscopically measured, we mean the detection of emission and/or absorption lines of the host galaxy.  3C\,279 is a flat-spectrum radio quasar (FSRQ) at $z=0.536$ \citep{3c279redshift} and was observed above 50 GeV by MAGIC during a flare \citep{albert}.  

FSRQs are blazars which show emission lines with an equivalent width of $\ge5$\AA\,, providing strong host-galaxy emission and absorption for spectroscopic distance measurements.  Conversely, BL Lacertae-type blazars display weak or no emission/absorption lines, and lack evidence for a Ca H$+$K break in their optical spectrum \citep{marcha, healey}.  With this definition, it is not surprising that there are VHE-detected BL Lac objects that lack definite spectroscopic redshift measurements.  Some notable VHE blazars without spectroscopic redshift are KUV\,00311-1938, PG\,1553+113, 3C\,66A, S5\,0716+714 and PKS\,1424+240 (PG\,1424+240).  

KUV\,00311-1938, detected at TeV energies by HESS \citep{HESS}, has a tentative redshift of 0.61 from the observation of five features in a poor signal-to-noise ratio (SN) spectrum \citep{piranomante}.   Follow-up spectroscopy with improved SN did not confirm any $z=0.61$ features, although an intervening Mg~II absorption system at $z=0.506$, not visible in the previous spectrum, was observed and sets a lower limit on the blazar distance \citep{pita}. 

Since BL Lac objects display bright and featureless spectra, strict lower limits on the redshifts can be inferred through observation of far-UV (FUV) absorption by the low-$z$ intergalactic medium (IGM).  This method has been applied to PG\,1553+113, 3C\,66A, and S5\,0716+714 showing lower limits of $z=$0.395, 0.3347 and 0.2314 (\cite{danforth2010,furniss,danforth2013}, respectively).  We apply this technique to the VHE-detected blazar PKS\,1424+240, constraining the blazar to $z\ge0.6035$, making it the most distant VHE-detected blazar thus far.   

\section{Observations and Far-UV Spectral Analysis}
PKS\,1424$+$240 was observed under a \textit{Hubble Space Telescope} (HST) program (12612, PI: Stocke) which uses flaring blazars as probes of intervening, weak IGM absorption.  PKS\,1424$+$240 is one of $\sim200$ objects monitored by a network of automated telescopes: the Katzman Automatic Imaging Telescope (KAIT, \cite{filippenko}), the ``NF/ Observatory" \citep{neely} and the Small and Moderate-Aperture Remote Telescope System (SMARTS, \cite{bonning}).  Optical photometry in April of 2012 indicated that PKS\,1424$+$240 was sufficiently bright to trigger a five-orbit HST/COS observation. 

The blazar was observed on 19 April 2012, with the medium resolution ($\Delta v\approx\rm 18~km~s^{-1}$), far-UV gratings G130M ($1135<\lambda<1450$ \AA, 6.4 ksec) and G160M ($1400<\lambda<1795$ \AA, 7.9 ksec).  The flux-calibrated, one-dimensional spectra were obtained from the Mikulski Archive for Space Telescopes and combined with the standard IDL procedures described in \citet{danforth2010}.   Additional analysis details are given in \cite{danforth2013prep}.  The combined data show a continuum flux level of $\sim1.4\times10^{-14}\rm~erg~cm^{-2}~s^{-1}~\AA^{-1}$ and a median SN per 7-pixel resolution element of $\sim19$ over the entire spectrum.   The spectrum was normalized with an iterative, spline-based procedure and an automated line-finding algorithm was used to locate $>4\sigma$ absorption features \citep{danforth2013prep}.  

We set a firm, observational lower limit to the source redshift by examining intervening absorbers.  We see absorption consistent with higher-order H~I Lyman absorption at $z\approx0.6$ (Figure~1).  The presence of absorption profiles consistent with Ly$\beta$ and Ly$\gamma$ at three distinct redshifts provides unambiguous line identifications of absorbing gas along the line of sight (Table~1).  The measurements summarized in Table~1 are made using Voigt-profile fits to the lines.  The significance levels are calculated from the observed equivalent widths and flux errors and are calculated via the methods described in \cite{Keeney12}.  The line identifications are bolstered in two cases by low-significance Ly$\delta$ detections.

In other cases where H~I absorption is used to constrain blazar distance, the {\em lack} of absorbers past a certain redshift is used to place a statistical upper limit. For PKS\,1424$+$240, the Ly$\alpha$ forest extends to the red end of the FUV detector ($\sim1800$ \AA, $z=0.47$).  Absorbers at higher redshift are detected through paired Ly$\beta$ and Ly$\gamma$ lines visible at $z<0.75$.  However this technique is less sensitive to H~I absorbers than detection via Ly$\alpha$ since both Ly$\beta$\ {\em and} Ly$\gamma$ lines must be detected to unambiguously identify an absorber and $(f\,\lambda)_{\rm Ly \gamma}$ is only $\sim5\%$ that of Ly$\alpha$.

No H~I absorbers are seen between the reddest system ($z=0.6035$) and the red edge of the detector in Ly$\beta$ ($z=0.75$) for a line-free path length $\Delta z=0.15$.  The spectral data quality is such that we should detect lines at $\log\,N_{\rm H~I}\ga14.0$ ($W_{\rm Ly \beta}\ga60$ m\AA, $W_{\rm Ly \gamma}\ga20$m\AA) at a $4\sigma$ level.  The frequency of lines of this strength or higher at low redshift is $d{\cal N}/dz\approx24$ \citep{danforth2008}, so we would expect ${\cal N}=3.5^{+3.1}_{-1.8}$ absorbers to be present if the source were at $z>0.75$.  Since no lines are seen, we can rule out $z>0.75$ at approximately $2\sigma$ confidence.  More detailed simulations may refine the redshift upper limit, but near-UV spectra and a direct search for Ly$\alpha$ lines at $z\sim1$ will be less ambiguous.  %We have applied for HST STIS observations which will enable a more reliable determination of a redshift upper limit.

The redshift lower limit is significantly higher than the conservative estimates of $z\ge0.06$ \citep{scarpa} and $z=0.23$ \citep{meisner}, derived from host-galaxy assumptions.  Notably, the redshift estimates in \cite{meisner} are in close agreement with other blazar distance limits derived from the observation of absorption systems.  The redshift limit for PKS\,1424+240 is also below the upper limits of $z=0.66$ and 1.19 derived from correcting VHE observations for EBL absorption to match the \textit{Fermi} Large Area Telescope (LAT) data in \cite{acciari1424} and \cite{yang}, respectively.  

\section{Absorption of Very High Energy Photons}
The energy- and redshift-dependent absorption of VHE gamma rays by the EBL can be estimated using the model-specific optical depth, $\tau(E,z)$, where the absorption-corrected (deabsorbed) flux, $F_{cor}$, is estimated from the observed flux, $F_{obs}$, using the relation $F_{cor} = F_{obs}\times e^{\tau(E,z)}$.   We investigate the deabsorbed VHE spectrum of PKS\,1424+240 at the redshift lower limit of $z=0.6035$, using three models to explore the effect of relatively low, medium, and high levels of EBL photon density.    The lowest density model used is from \cite{gilmore2012}, which estimates a $z\sim0$ EBL spectral energy distribution nearing the required lower limits on the EBL set by galaxy counts.  We use the \cite{dominguez} model to estimate VHE absorption by an intermediate level EBL density and the \cite{finke} model to represent a relatively high level of EBL density.%, which is a model derived from templates describing star formation and dust reemission. 

\subsection{Constraining the Opacity of the EBL}
The blazar VHE spectral index can be used to estimate the EBL spectral properties under the assumption that the intrinsic spectrum, characterized by the power-law $dN/dE \propto E^{-\Gamma}$, cannot be harder than $\Gamma=1.5$, as described in \cite{aharonian2006}.  The limit is physically motivated by the shock acceleration paradigm, where the hardest index obtained for the accelerated leptons is 1.5.  %Moreover, this same slope results for gamma-ray production via proton interaction with the ambient plasma (see \cite{malkov2001, blandford}).   Slightly harder emission spectra ($\Gamma=1.25$) are possible for emission occurring entirely in the Thomson regime and lacking efficient radiative cooling, although this scenario is unlikely at high energies \citep{aharonian2006}. 

%Notably, the hardest blazar spectral index measured by \textit{Fermi} LAT agrees with this limit \citep{2fgl}.  
Notably, the hardest blazar spectral index measured by the LAT has an index of 1.1 but is in statistical agreement with the theoretical limit \citep{ackermann2011,2fgl}.  Since the LAT is most sensitive to photons at energies where EBL absorption is negligible, the indices derived from LAT observations reflect the intrinsically emitted gamma-ray spectrum.  A more conservative limit equal to the LAT measured index can be placed %on the absorption-corrected VHE spectral index
under the assumption that blazars do not harden with increasing energy.  For PKS\,1424+240, the contemporaneous LAT-measured index is 1.80$\pm$0.07 \citep{acciari1424} (Figure~3).

%Since there may be two independent sources of gamma-ray absorption at work for the blazar VHE photons, namely from the EBL and from a BLR, the expected BLR opacity as a function of energy is an important consideration.  As pointed out by \cite{aharonianKC}, the intrinsic absorption of gamma-rays within a BLR can harden intrinsically emitted spectra.  More specifically, the BLR-absorbed spectrum could be harder than the intrinsically emitted one if the optial depth, $\tau_{BLR}(\rm E)$, decreases with energy, in turn breaking the $\Gamma=1.5$ limit set by standard emission scenarios.  If, instead, the optical depth remains constant, or increases with energy, the spectral constraints on the opacity of the EBL remain valid.  \cite{tavecchio} show that the optical depth remains constant in the 100-500 GeV energy range where the PKS\,1424+240 VHE spectrum is measured by VERITAS through the modeling of the expected optical through ultraviolet radiation field within a BLR.

%The VERITAS observed spectrum of PKS\,1424+240 is described with a differential power-law with spectral index $\Gamma=3.8\pm0.5_{stat}$ \citep{acciari1424}.  
A power-law fit can be applied to the absorption-corrected points for the redshift lower limit of $z=0.6035$, as summarized in Table 2 for each of the EBL models.  %These fitted indices can be compared to the contemporaneous \textit{Fermi} LAT measured spectrum of 1.77$\pm$0.02.  
The fitted indices for the deabsorbed spectra using the relatively low and medium EBL models (\cite{gilmore2012} and \cite{dominguez}, respectively) are well within the $\Gamma$=1.5 and $\Gamma=1.80\pm0.07$ limits.  However, the $\Gamma=0.6\pm0.8$ index resulting from deabsorption with the relatively more opaque model from \cite{finke} is below, but still consistent with, either of the expectations.  Improved gamma-ray observations may reveal that this model is too dense.   Other explanations, such as time-dependent stochastic accelerated inverse-Compton scenarios \citep{lefa2011a,lefa2011b} and internal gamma-gamma absorption \citep{aharonian2008}, can also account for unusually hard VHE emission spectra.

\subsection{The Gamma-ray Horizon}
Intergalactic gamma-ray opacity due to the EBL has direct consequences for the estimation of the intrinsic gamma-ray spectra of extragalactic VHE targets, with the source emitted flux being suppressed by $e^{-\tau(E,z)}$.  This energy- and redshift-dependent flux suppression requires sources to be exponentially brighter at larger distances in order to be detectable at VHE.% by current generation instruments such as VERITAS, MAGIC, or HESS.

The opacities probed through the VHE observation of blazars with redshift information provides insight into the possibility of a pair-production anomaly, as investigated in \cite{hornsmeyer}.  In the seven VHE blazar spectra which probe opacities in the range 1$\le\tau\le2$, an upturn of the absorption-corrected spectra is apparent with a significance of 4.2$\sigma$ at the $\tau\ge2$ transition energy.  Due to the different energies of the $\tau\ge2$ transition for the blazars studied, source-intrinsic features are an unlikely explanation.% for the observed effect.  
This study was limited by the number of VHE blazars probing opacities $\tau\ge2$ with known redshift.  PKS\,1424+240 can now be included in the study, expanding the limited opacity parameter space available.% to infer the apparent pair-production anomaly between EBL and VHE photons.

Before the determination of the redshift lower limit of PKS\,1424+240, the highest sampled gamma-ray optical depth probed was associated with the detection of 3C\,279 during an elevated state, probing a $\tau({\rm E}=475\, {\rm GeV}, z=0.536)=3.2$, 3.9 or 4.3 when estimated with the \cite{gilmore2012,dominguez,finke} models, respectively.   Now, with PKS\,1424+240, the highest opacity sampled is between 4.1 and 5.3 (if estimated with the \cite{gilmore2012} and \cite{finke} models, respectively.)%, while with the less opaque \cite{gilmore2012} model the opacity reaches $\tau=4.1$, showing the steady state of PKS\,1424+240 to be significantly brighter than 3C\,279 in a flaring state.  

We show the highest opacity sampled by the VHE detection of 3C\,279 compared to the opacity probed with the detection of PKS\,1424+240 in Figure~2, as derived for the \cite{dominguez} model.  The spectral points are plotted along with the $\tau(E,z)=1-5$ horizons.  For reference, we also show published spectral points of every VHE-detected blazar with a spectroscopically measured redshift above 0.2 (four sources, as compared to the more than 30 VHE blazars with $z<0.2$).  Since the redshift of PKS\,1424+240 represents a lower limit, the maximum opacity sampled by the VHE detection is illustrated with a right-pointing arrow.

\section{Absorption-Corrected Gamma-ray Emission}
%Without information about a VHE-detected blazar redshift, only the observed broadband emission can be modeled, as done with a synchrotron self-Compton emission scenario in \cite{acciari1424} for redshift values between 0.05 and 0.7.  
The observed gamma-ray peak is reproduced from \cite{acciari1424} in Figure~3, illustrating the contemporaneous LAT and VERITAS data with black squares and circles, respectively.  Additionally, we correct the observed VHE spectrum for EBL absorption at the minimum redshift of $z=0.6035$ using the intermediate model \citep{dominguez}, illustrated by grey circular points.  Absorption of the LAT data is negligible.%are at low enough energy that the absorption by the EBL is negligible.  

The LAT and VERITAS gamma-ray observations extend from 0.5 to 500 GeV.  This full range, when corrected for EBL absorption at $z=0.6035$, is not well fit with a power law (dashed blue line), while a curved log parabola provides an improved fit (dotted blue line).  We note than neither the power-law nor the log-parabolic fit represents the data well.  %Although only the intermediate absorption-corrected VHE flux values from \cite{dominguez} are shown in Figure 3, 
The fit results for each EBL model are summarized in Table~2.  Both the power-law and log-parabolic fits find indices of $\Gamma\sim2$, with a significantly curved log-parabolic fit.   Notably, the highest energy VERITAS point at 500 GeV appears to deviate from the fits, matching the LAT extrapolated power-law (red dashed line).  This deviation is not expected by standard blazar emission models, but is only about two standard deviations at the redshift lower limit.

A break between the LAT and VERITAS absorption-corrected data is apparent.  This discrepancy is not likely an issue of instrumental cross-calibration, as agreement between VERITAS and LAT observations have been found for other contemporaneous blazar observations (e.g., \cite{rxj0648,rbs0413}).  A portion of this feature may be due to a small level of undetected variability.  Although short intervals of variability are difficult to rule out, no long-term variability is detected \citep{acciari1424}, making it unlikely that the spectral feature between the LAT and VERITAS instruments is due to variability alone.

Since $z=0.6035$ is a lower limit, it is conceivable that the discontinuity between the LAT and VERITAS data may, in fact, be an unphysical effect arising from the incomplete correction for absorption by the EBL.  The first differential flux point of the VHE spectrum at 150 GeV cannot be made to match the LAT extrapolated spectrum without deabsorbing the flux for a redshift of 1.2 (blue stars in Figure~3).  This deabsorbed spectrum is shown for the \cite{dominguez} model, but is representative of the required distances of $z=1.5$ and $z=1.0$ when corrected with the \cite{gilmore2012} and \cite{finke} models, respectively.  

The $z=1.2$ corrected VHE spectrum results in a \textit{rising} slope, i.e. with an index $\Gamma=-2.5\pm1.0$ when fit with a differential power law.  A VHE spectrum with a rising power law is difficult to produce in even the most extreme emission scenarios.  Although this redshift value is still in agreement with the redshift upper limit set by \cite{yang}, we interpret the unphysical VHE spectrum as evidence that the blazar does not reside at this distance.

\subsection{Possible Signature of Intrinsic Gamma-ray Absorption}
Assuming that the blazar resides near $z=0.6035$, the apparent discontinuity between the LAT and VERITAS energy ranges may be due to gamma-ray absorption in the vicinity of the blazar.  It has been shown that absorption of gamma-rays by a broad line region (BLR) can produce broken power-law spectra in bright LAT-detected blazars \citep{poutanen}.  However, this type of absorption is not immediately expected for an intermediate/high-synchrotron-peaked source such as PKS\,1424+240, expected to exhibit a clean radiation environment \citep{bottcher, ghisellini}.  

The absorption-corrected VHE point at 500 GeV matches the LAT power-law extrapolation, with a distinct mismatch to the 100 through 400 GeV points.  A source of gamma-ray opacity that is only sensitive to photons between 100 GeV and $\sim$400 GeV is difficult to explain with an ion continuum such as that present in a BLR.  It has been shown that the optical depth of a BLR containing UV-continuum and ionization lines produces a constant optical depth from tens of GeV to beyond 30 TeV \citep{tavecchio}.   Since PKS\,1424+240 is not expected to harbor a BLR, it is perhaps more likely that the EBL model is slightly overcorrecting for the photon absorption around 500 GeV.  The observed hard spectrum might also be explained by secondary gamma rays produced in cosmic-ray interactions along the line of sight \citep{essey,aharonian2013}.%An overcorrection could be misinterpreted as an absorption feature, causing the observed flux to deviate from a power law.  The absorption-corrected VHE upper limit at 750 GeV is high enough to allow for a continuation of this power-law beyond the 500 GeV point, but nothing can be said with certainty without VHE measurements at higher energies.  

\section{Conclusions} 
We present the strict redshift lower limit of $z\ge0.6035$ for PKS\,1424+240, set by the detection of Ly$\beta$ and $\gamma$ lines from intervening hydrogen clouds.  This lower limit makes PKS\,1424+240 the most distant VHE-detected source.  At this distance, VHE observations of the source out to energies of 500 GeV probe gamma-ray opacities of up to $\tau\sim$5.

An investigation of possible constraint on the opacity of the EBL shows that the absorption-corrected power-law fits do not lie significantly outside of the standard spectral limitations. However, deabsorption with the \cite{finke} model produces the hardest intrinsic VHE spectrum.  If the blazar resides at a redshift beyond the lower limit, the deabsorbed indices may become constraining to even the lowest level EBL model.  

This redshift information allows the investigation of the EBL absorption-corrected gamma-ray emission.  Correcting the VHE spectrum for the minimum redshift of $z=0.6035$ shows an unexpected spectral shape.  The elevated flux around 500 GeV, while not high significance, may be due to an overcorrection of absorption by the EBL model.  This feature is still present when using the \cite{gilmore2012} model (open circles in Figure~3), which predicts a low EBL density and therefore might be evidence for a pair-production anomaly, such as might occur if VHE photons mixed with axion-like particles.

The spectral feature occurring between the LAT and VERITAS energy bands cannot be reasonably removed by correcting for additional absorption due to a larger distance.  Instead, deabsorption for a higher redshift to match the LAT and VERITAS fluxes results in a non-physical spectrum and an extreme distance of $z=1.2$.  It is possible, although unlikely, that the EBL absorption-corrected gamma-ray spectrum of PKS\,1424+240 is exhibiting gamma-ray absorption within an intrinsic photon field such as a BLR.   As the most distant VHE BL Lac object, probing historically high values of gamma-ray opacity, this source requires additional gamma ray observations and a tighter constraint on the redshift to further investigate the intrinsic emission.

\acknowledgments

Support for programs HST-GO-12621, 13008 and 12863 and for Hubble Fellow M.F. (grant HF-51305.01-A) was provided by NASA awarded through the Space Telescope Science Institute, operated by the Association of Universities for Research in Astronomy,  Inc., for NASA, under contract NAS 5-26555.  Additional support came from the National Science Foundation award PHY-0970134. C. D. was supported by NASA grants NNX08AC146 and NAS5-98043.  A.V.F. is grateful for the NASA/Fermi grants GO-31089 and NNX12AF12GA, NSF grant AST-1211916, the Christopher R. Redlich Fund, and the TABASGO Foundation. KAIT operation is made possible by donations from Sun Microsystems, Inc., the Hewlett-Packard Company, AutoScope Corporation, Lick Observatory, the NSF, the University of California, the Sylvia \& Jim Katzman Foundation and the TABASGO Foundation. We thank the late Weidong Li for setting up the KAIT blazar monitoring program and S. Bradley Cenko for retrieving the KAIT data.

{\it Facilities:} \facility{VERITAS}, \facility{HST (COS)}, \facility{Fermi (LAT)}.

\clearpage

\begin{deluxetable}{llcccc}
  \tabletypesize{\scriptsize}
  \tablecolumns{6}
  \tablewidth{0pt}
  \tablecaption{Intervening Lyman Absorption Lines}
  \tablehead{\colhead{Line}   &
            \colhead{$\lambda_{\rm obs}$ (\AA)}    &
            \colhead{$W_{\rm r}$ (m\AA)\tablenotemark{a}}    &
            \colhead{$b\rm~(km\,s^{-1})$\tablenotemark{b}}     &
            \colhead{$\log\,N$ (cm$^{-2}$)\tablenotemark{c}} &
            \colhead{S.L. ($\sigma$)\tablenotemark{d}}   
  }
         \startdata
  \sidehead{$z=0.5838$ system}
    Ly$\beta$  & 1624.6 &$ 89\pm29$ &$ 47\pm7$&$ 14.18\pm0.12$&  6 \\
    Ly$\gamma$ & 1540.2 &$ 73\pm 5$ &$ 55\pm8$&$ 14.52\pm0.04$&  5\\
  \sidehead{$z=0.5960$ system}
    Ly$\beta$  & 1637.1 &$180\pm 1$ &$ 44\pm3$&$ 14.52\pm0.02$& 11 \\
    Ly$\gamma$ & 1552.1 &$ 60\pm33$ &$ 37\pm5$&$ 14.44\pm0.19$&  5 \\
    Ly$\delta$ & 1515.7 &$ 19\pm 6$ &$ 31\pm5$&$ 14.3\pm0.2$& $\sim2$ \\
  \sidehead{$z=0.6035$ system}
    Ly$\beta$  & 1644.8 &$ 70\pm11$ &$ 14\pm2$&$ 14.15\pm0.06$&  7 \\
    Ly$\gamma$ & 1559.5 &$ 36\pm10$ &$ 15\pm3$&$ 14.25\pm0.08$&  4 \\
    Ly$\delta$ & 1522.9 &    $\sim$13    &   $\sim$10   &$ 14.1 \pm0.3 $& $\sim2$ \\
  \enddata
  
  \tablenotetext{a}{Rest-frame equivalent width.}
  \tablenotetext{b}{Doppler parameter.}
  \tablenotetext{c}{Corresponding column density.}
  \tablenotetext{d}{\,Line significance (according to \cite{Keeney12}).}
\end{deluxetable}

\begin{deluxetable}{lccccccc}
\tabletypesize{\scriptsize}
\tablecaption{Results for the power-law fit to the absorption-corrected VHE points from \cite{acciari1424} for $z=0.6035$.  Additionally, we show the fits to the full range of data (0.5-500 GeV) for a power law ($dN/dE\propto(E/E_o)^{-\Gamma}$) and log parabola ($dN/dE\propto(E/E_o)^{-\Gamma-\beta {\rm Log}(E)}$). The fits for the \cite{dominguez} model are shown in Figure~3 in blue dashed and dotted lines, respectively.}
\tablewidth{0pt}
\tablehead{
  \colhead{EBL} &  \colhead{VHE Range} &\colhead{$\tilde{\chi}^2$} &\colhead{Full Range} &\colhead{$\tilde{\chi}^2$} & \colhead{Full Range} &\colhead{Full Range} &\colhead{$\tilde{\chi}^2$}\\
  \colhead{Model}&  \colhead{Deabsorbed}&\colhead{(4 DOF)}& \colhead{Deabsorbed} &\colhead{(9 DOF)}& \colhead{Deabsorbed} &\colhead{Deabsorbed}& \colhead{(8 DOF)}\\
  \colhead{Used}&  \colhead{Power-law}&\colhead{} & \colhead{Power-law}&\colhead{}& \colhead{Log-parabola} & \colhead{Log-parabola}& \colhead{} \\  
  \colhead{}&  \colhead{Index}&\colhead{}& \colhead{Index}&\colhead{} &\colhead{Index}  & \colhead{Curvature}&\colhead{}\\  
  \colhead{}&  \colhead{$\Gamma$} &\colhead{}& \colhead{$\Gamma$} &\colhead{}& \colhead{$\Gamma$}  & \colhead{$\beta$}&\colhead{}\\}   
   
 \startdata
\cite{gilmore2012} &1.5$\pm$0.8&0.85&2.07$\pm$0.03&2.3&2.04$\pm$0.04  & 0.10$\pm$0.03&1.4\\
\cite{dominguez}   &1.0$\pm$0.7&0.88& 2.01$\pm$0.03&2.0& 1.99$\pm$0.04 &0.08$\pm$0.03 &1.5\\
\cite{finke}              &0.6$\pm$0.8&0.83&1.97$\pm$0.04 &1.9& 1.96$\pm$0.04  &0.07$\pm$0.03 &1.5\\
\enddata

\end{deluxetable}

\begin{figure}
\includegraphics[scale=0.9]{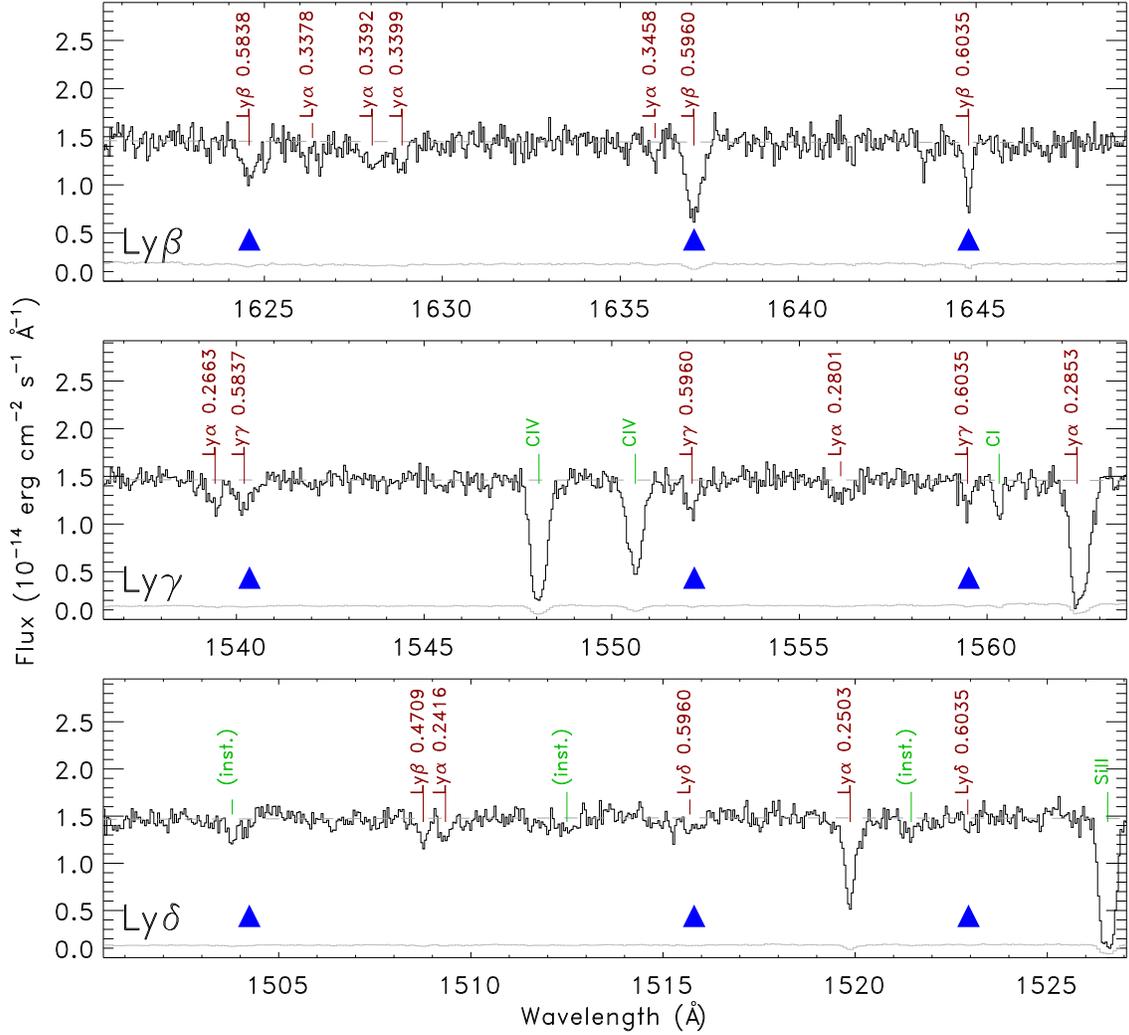}
  \caption{COS spectra show intervening absorption systems at $z=0.5838$, $z=0.5960$, and $z=0.6035$ (arrows) toward PKS\,1424+240 in Ly$\beta$ (1025.72\AA), Ly$\gamma$ (972.54 \AA), and Ly$\delta$ (949.74 \AA).  COS flux and error (grey) vectors are binned by four pixels (half a resolution element).  Continuum fit is shown with dashed line.  Other intervening absorption is identified with species and redshift in red. Ly$\alpha$ absorption systems are observed to the edge of the detector (1800$\AA\,$, corresponding to z$_{Ly\alpha}<0.47$).  Galactic ($v\approx0$) absorption and instrumental features are labeled in green.  Ly$\alpha$, $\beta$ and $\gamma$ features are detected at  $\ge3\sigma$ and will be further discussed in \cite{danforth2013prep}.
\label{fig1}}
\end{figure}

\begin{figure}
\includegraphics[scale=0.8]{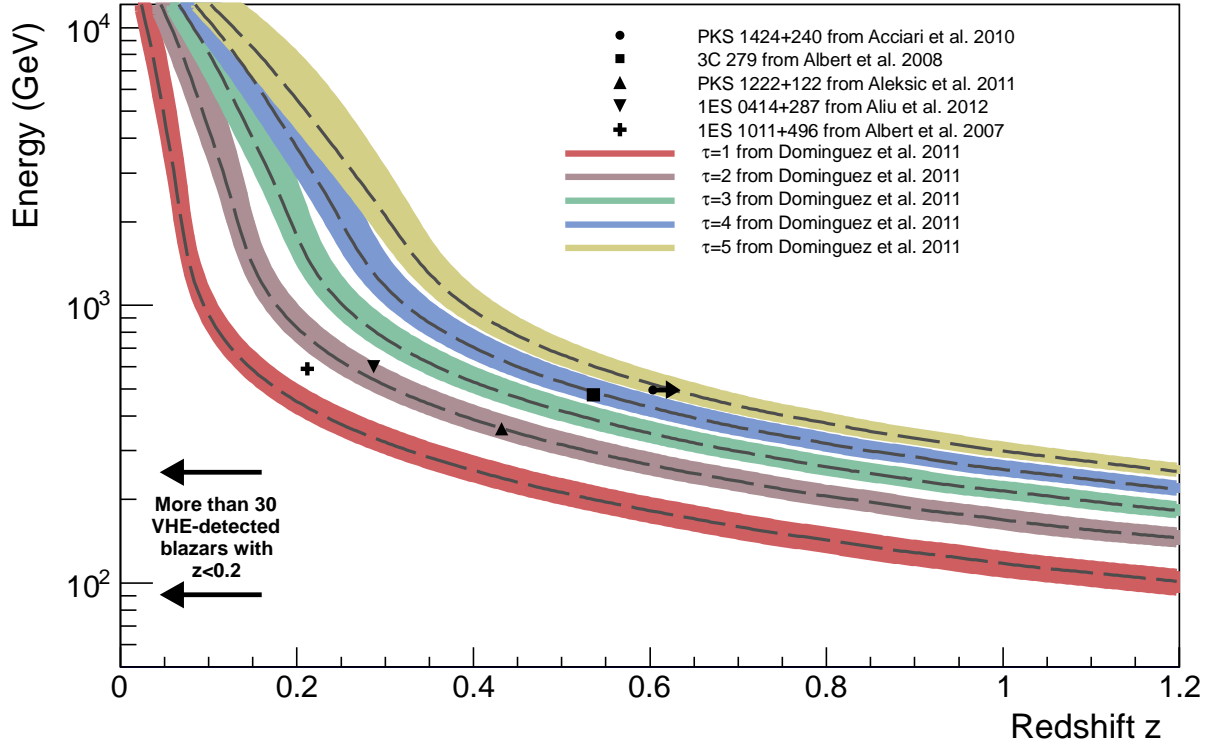}
\caption{The highest energy points of the VHE-detected blazars with published VHE data and spectroscopic redshifts beyond 0.2.  %There are only five spectroscopically measured blazars with published VHE data which reside beyond $z=0.2$, as compared to the more than 30 at $z<0.2$.  
The close proximity of VHE blazars is a result of the gamma-ray opacity of the Universe.  The $\tau=1-5$ gamma-ray horizon contours are shown as bands, including model errors, representing the energy and redshift dependent $e^{-\tau}$ suppression of the VHE flux for extragalactic sources as calculated from \cite{dominguez}.  The VHE detection of PKS\,1424+240 is shown with a rightward arrow, indicating the redshift is a lower limit. \label{fig2}}  
%\caption{VERITAS spectral measurements of PKS\,1424+240 from \cite{acciari1424} at the redshift lower limit of $z=0.6035$, shown with respect to the MAGIC spectral measurements of FSRQ 3C\,279 at $z=0.536$ from \cite{albert} in relation to the $\tau=1-5$ gamma-ray horizons, as estimated by the EBL model \cite{dominguez}.  The firm redshift lower limit places PKS\,1424+240 as the most distant VHE emitting source detected to date, where past VHE spectral measurements probe unprecedented gamma-ray opacity values of $\tau\sim5$.\label{fig2}}
\end{figure}

\begin{figure}
\includegraphics[scale=0.8]{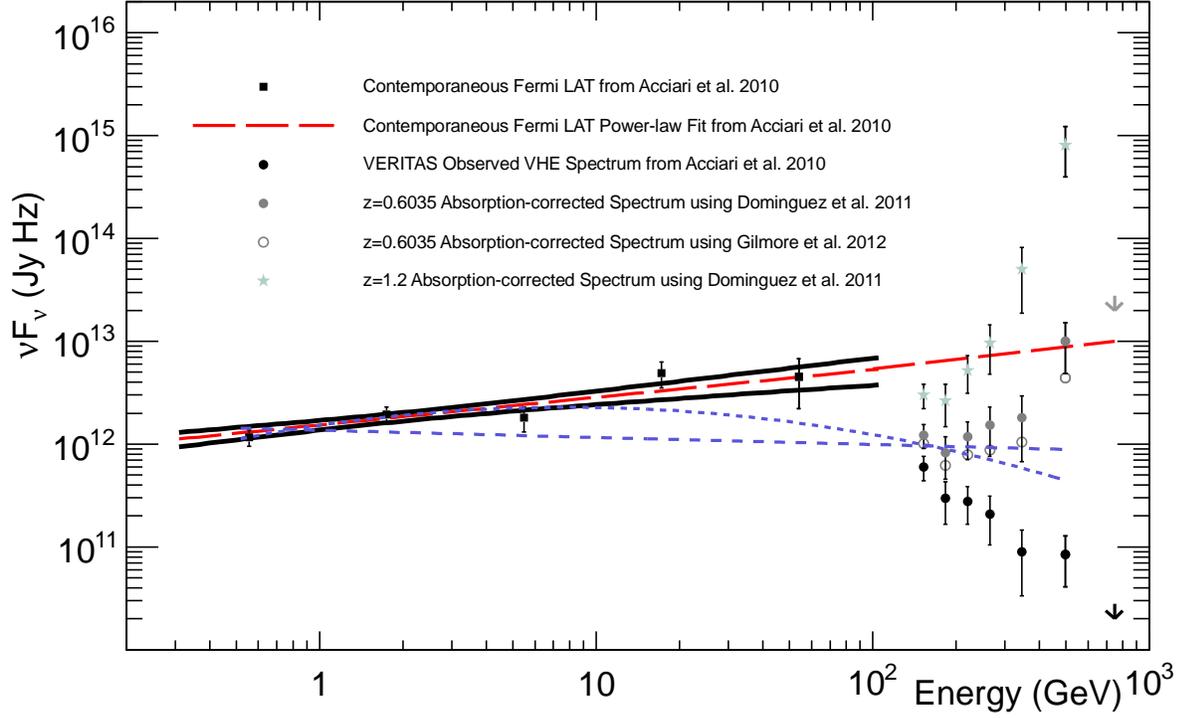}
\caption{The gamma-ray peak of the spectral energy distribution of PKS\,1424+240, with LAT (squares and power-law fit contour) and VERITAS observations (black circles) taken from \cite{acciari1424}.  An upper limit at 750 GeV is shown with a downward pointing arrow.  The LAT data have been selected to be contemporaneous with the VERITAS observations. The absorption-corrected VHE spectrum is shown with the grey circles, using opacities from the \cite{dominguez} EBL model.  For reference, the absorption-corrected points using the \cite{gilmore2012} model are shown in open circles, with errors (not drawn) similar to those shown for the \cite{dominguez} deabsorbed points.  The LAT power-law fit has been extrapolated up to VHE (dashed red line).  Power-law and log-parabolic fits to the full range (0.5-500 GeV) are shown in the blue dashed and dotted lines, respectively, with fitting results in Table 2.  To bring the first absorption-corrected VERITAS spectral point to match the LAT observed spectrum, the blazar needs to be corrected for absorption expected for $z\approx1.2$, shown by blue stars (the upper limit for this deabsorption is off-scale). % This higher redshift would require a significant hardening of the spectrum beyond $\sim$200 GeV, with an unphysical rising power-law index, a scenario not easily described by traditional acceleration mechanisms. 
\label{fig3}}
\end{figure}

\end{document}